\documentclass[pra,twocolumn,superscriptaddress,showpacs,showkeys]{revtex4}
\usepackage{amsmath}
\usepackage{graphicx}
\usepackage{dcolumn}
\usepackage{bm}
\usepackage{amssymb}
\usepackage{epsfig}

\begin{document}
\bibliographystyle{apsrev}


\title{Mesoscopic Cavity Quantum Electrodynamics with Quantum Dots}

\author{L. Childress}
\affiliation{Department of Physics, Harvard University, Cambridge,
Massachusetts, 02138}

\author{A. S. S\o rensen}
\affiliation{Department of Physics, Harvard University, Cambridge,
Massachusetts, 02138}
\affiliation{ITAMP, Harvard-Smithsonian Center for Astrophysics, Cambridge,
Massachusetts, 02138}

\author{M. D. Lukin}
\affiliation{Department of Physics, Harvard University, Cambridge,
Massachusetts, 02138}
\affiliation{ITAMP, Harvard-Smithsonian Center for Astrophysics, Cambridge,
Massachusetts, 02138}

\date{\today}

\begin{abstract}
 We describe an electrodynamic mechanism for coherent, quantum mechanical coupling between spacially separated quantum dots on a microchip.  The technique is based on capacitive interactions between the electron charge and a superconducting transmission line resonator, and is closely related  to atomic cavity quantum electrodynamics.  We investigate several potential applications of this technique which have varying degrees of complexity.   In particular, we demonstrate that this mechanism allows design and investigation of an on-chip double-dot microscopic maser.   Moreover, the interaction may be extended to couple spatially separated electron spin states while only virtually populating fast-decaying superpositions of charge states.   This represents an effective, controllable long-range interaction, which may facilitate implementation of quantum information processing  with  electron spin qubits and potentially allow coupling to other quantum systems such as atomic or superconducting qubits.
\end{abstract}

\maketitle
\section{Introduction}
Recent progress in quantum control of atoms, ions, and photons has spurred interest in developing architectures for quantum information processing \cite{IonsAndAtoms}.   An intriguing question is whether similar techniques can be extended to control quantum properties of  ``artificial atoms'' in a condensed matter environment.  These tiny solid state devices, e.g. flux lines threading a superconducting loop, charges in cooper pair boxes, and single electron spins,  exhibit quantum mechanical properties which can be manipulated by external currents and voltages \cite{JJ, GLoss}.   To realize their potential as highly tunable qubits, they must interact at a rate faster than the decoherence caused by the complex and noisy environment they inhabit.   Strong coupling between qubits is therefore essential to achieve a high degree of control over quantum dynamics.  For most systems a mechanism achieving the required coupling strength has only been proposed for nearby qubits \cite{DiVi98}, thus limiting the spatial extent of controllable interactions. 

We describe a technique for coupling mesoscopic systems that can be millimeters apart.  In our proposal, a strong interaction is obtained by linking charge qubits to quantized voltage oscillations in a transmission line resonator.   We show that the capacitive coupling between charge degrees of freedom of the mesoscopic system and the superconducting resonator is formally analogous to cavity quantum electrodynamics (cavity QED) in atomic physics \cite{CQEDRMP}.  Such an interaction may be used for controllable coupling of distant mesoscopic qubits, thereby facilitating scalable quantum computing architectures.   Furthermore, we have recently shown \cite{Anders} that by combining the present approach with atom trapping above the resonator, these techniques may allow coupling between solid state and atomic qubits, thus opening a new avenue for quantum information research.   

These new opportunities for qubit manipulation using microwave photons are made possible by the excellent coherence properties of high quality factor superconducting microstrip resonators originally developed for photon detection \cite{Resonators}.   With observed Q-factors exceeding $10^6$ at 10 GHz, such resonators could permit on-chip storage of a microwave photon for more than $10\ \mu$s.  Moreover, in contrast to the microwave cavities used in atomic cavity QED \cite{Haroche, Walther}, these 1D transmission line resonators have mode volumes far smaller than a cubic wavelength, allowing a significantly stronger coupling to resonator modes.   This combination of long coherence time and strong  coupling makes microstrip resonators a promising technology for quantum manipulation.

In this paper we outline several intriguing avenues for applications of these resonators in the context of quantum dot research.  We first discuss a mechanism for strong coupling  between spacially separated charge states in a mesoscopic system.   We show in particular that the use of quantum dots may allow construction of novel quantum devices such as an on-chip double-dot microscopic maser \cite{MandS, Haroche, Walther}.  Our discussion pertains to lithographically defined lateral double quantum dots \cite{DoubleQD}, although the resonator coupling mechanism would apply equally to other mesoscopic systems.  Indeed, similar ideas for resonator mediated interactions have been developed  in the context of superconducting qubits \cite{JJRes}.  Specifically, the strong-coupling mechanism analogous to that presented here has been proposed independently in \cite{Girvin}.  

Although coherence properties of charge states in quantum dots are likely worse than those of superconducting systems, quantum dots have two potential advantages:  they are highly tunable, and the electrons they confine are not paired, allowing access to the electron spin.    We show that the more stable spin degree of freedom may be accessed using techniques for quantum coherent manipulation initially developed in atomic physics \cite{RAP, CQEDQC}.    In analogy to the use of Raman transitions in cavity quantum electrodynamics, the electron spin states can be coupled via a virtual charge state transition.    Since the spin decoherence rate is far slower than the charge decoherence rate, the losses can thereby be greatly reduced.  Finally, we explicitly address the effect of the resonator on radiative contributions to qubit decoherence, and demonstrate that the latter can be greatly suppressed.

Before proceeding, we also note important earlier work on cavity quantum electrodynamics with quantum dots in the optical regime \cite{Imamoglu}, which differs qualitatively from the ideas discussed here.

\section{Cavity QED with Charge States}

\subsection{The resonator-double dot interaction}

\begin{figure}[htbp]
\centerline {
\includegraphics[width=3.5 in]{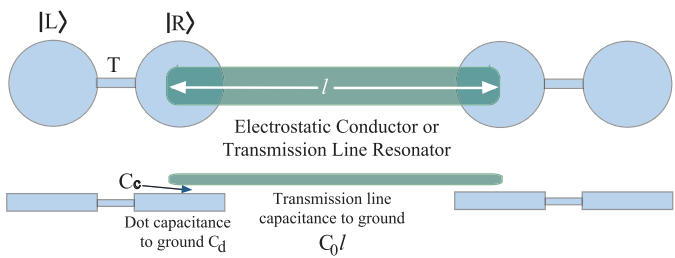}
}
\caption{\it{ Two double dots coupled by a conductor with capacitances as described in the text. Note that a transmission line resonator requires a nearby ground plane (not shown) to shield the system from radiative losses.  } } 
\label{interdot}
\end{figure}

We consider a single electron shared between two adjacent quantum dots whose energy eigenstates are tuned close to resonance with the fundamental mode of a nearby superconducting transmission line segment.   The electron can occupy the left or right dot, states $|L\rangle$ and $|R\rangle$ respectively, and it tunnels between the dots at rate $T$ (see Figure~\ref{interdot}).    A capacitive coupling $C_c$ between the right dot and the resonator causes the electron charge state to interact with excitations in the transmission line.   We assume that the dot is much smaller than the wavelength of the resonator excitation, so the interaction strength may be derived from the electrostatic potential energy  of the system, $\hat{H}_{\mathrm{Int}} =  e \hat{V} \upsilon |R\rangle\langle R|$, where $e$ is the electron charge, $\hat{V}$ is the voltage on the resonator near the right dot, $\upsilon =C_c/(C_c+C_d)$, and $C_d$ is the capacitance to ground of the right dot.

A more useful form of the interaction Hamiltonian is found by rewriting this energy in a different basis.   First,  we express the left and right dot states $|L\rangle$ and $|R\rangle$ in terms of the  double dot eigenstates.  If the two dots are tunnel coupled with matrix element $T$ and have a potential energy difference of $\Delta$, then the double dot eigenstates are given  by
\begin{eqnarray}
|+\rangle &=&\sin{\phi}|L\rangle + \cos{\phi}|R\rangle \\
|-\rangle &=&   \cos{\phi}|L\rangle-\sin{\phi}|R\rangle,
\end{eqnarray}
where $\tan{\phi} = - 2 T/ (\Omega +\Delta)$ and  $ \Omega =  \sqrt{4 T^2 + \Delta^2}$ is the splitting in frequency between the eigenstates $|+\rangle$ and $|-\rangle$.   For notational simplicity, we represent the electron charge state in terms of Pauli spin matrices by defining raising and lowering operators $\Sigma_+ = \Sigma_-^{\dagger} = |+\rangle\langle-|$,  so that $\Sigma_x = \Sigma_+ + \Sigma_-$, and so on. 
 
Next, we express the voltage as an operator.    A transmission line segment of length $l$, capacitance per unit length $C_0$ and characteristic impedance $Z_0$ has allowed wavevectors  $k_n = \frac{(n+1) \pi}{l}$ and  frequencies $\omega_n= \frac{k_n}{C_0 Z_0}$.    Canonical quantization of the transmission line Hamiltonian allows the voltage to be written in terms of creation and annihilation operators $\{\hat{a}^{\dagger}_n, \hat{a}_n\}$ for the modes $k_n$ of the resonator.  These excitations may be interpreted as microwave photons.   With substitution of the quantized voltage at the end of the resonator, 
\begin{equation}
\hat{V} = \sum_n{ \sqrt{\frac{\hbar \omega_n}{l C_0}}\left(\hat{a}_n +\hat{a}_n^{\dagger}\right)},
\end{equation}
the full Hamiltonian becomes
\begin{equation}
\hat{H} = \frac{\hbar \Omega}{2}\Sigma_z +  \sum_{n} \hbar \omega_n \hat{a}_n^{\dagger}\hat{a}_n +\hbar (g_z^{(n)} \Sigma_z + g_x^{(n)}\Sigma_x) (\hat{a}_n^{\dagger}+\hat{a}_n). 
\label{SB}
\end{equation}
The coupling constants,  $g_x^{(n)} = g_0 (T/\Omega)\sqrt{\omega_n/\omega_0}$ and $ g_z^{(n)} =  g_0 (\Delta/2\Omega)\sqrt{\omega_n/\omega_0}$ scale as an overall coupling strength
\begin{equation}
g_0  = \omega_0 \upsilon \sqrt{\frac{2 Z_0}{R_Q}}
\end{equation}
where $R_Q =\frac{h}{e^2}\approx 26$ kOhm is the resistance quantum.  

For $g_0 \ll \omega_0$ (which is guaranteed for low-impedance resonators $Z_0 \ll R_Q$) the Hamiltonian of Eq.~(\ref{SB}) may be further simplified by neglecting all terms which do not conserve energy:   If the dot is near resonance with the fundamental frequency, $\omega_0 = \frac{\pi}{l Z_0 C_0} \approx \Omega$, we may neglect all other modes, and in the rotating wave approximation (RWA) the Hamiltonian reduces to
\begin{equation}
\hat{H} \approx \hat{H}_{\mathrm{JC}} = \frac{\hbar \Omega}{2} \Sigma_z  +  \hbar \omega_0 \hat{a}^{\dagger}\hat{a}  +  \hbar g_0\frac{T}{\Omega}(\hat{a}^{\dagger} \Sigma_- + \hat{a}\Sigma_+).
\label{JC}
\end{equation}

The Jaynes-Cummings interaction described by Eq.~(\ref{JC}) furnishes a direct analogy to a two-level atom coupled to a single mode field.  A novel feature of the double dot system is that the parameters $\Omega(t)$ and $T(t)$ can be adjusted on fast time scales by varying the voltage applied to the metallic gates defining the quantum dots.  Consequently the detuning $\Omega(t) -\omega_0$ and the effective coupling constants $\propto \frac{T(t)}{\Omega(t)}$ may be controlled independently for double dots on either end of the resonator, each of which interacts with the resonator via the coupling described by Eq.~(\ref{JC}).  

To illustrate the strength of the resonator mediated interaction rate $T g_0/\Omega$, we compare it with a static interaction achieved by capacitively coupling spatially separated double dots through a conductor.  By calculating the change in electrostatic energy of an electron in one double dot due to shifting the electron between dots in the second double dot, we find that the electrostatic interaction energy is $\Delta E \approx \upsilon^2 e^2 / (C_0 l)$.   If the nonresonant conductor and the transmission line have the same length $l$ and capacitance to ground $C_0$,  the two interaction rates may be compared directly: 
\begin{equation}
\hbar g_0=  \frac{1}{\upsilon}\Delta E\sqrt{\frac{R_{Q}}{2 Z_0}}.
\end{equation}
Typically, $Z_0 = 50 \ \mathrm{Ohms} \ll R_Q$ and  careful fabrication permits a strong coupling capacitance, with $\upsilon \approx 0.28$ \cite{Chan}, so that $\hbar g_0 \approx 57 \Delta E$.  Hence a much stronger coupling can be achieved using the resonant interaction.
For example, a  wavelength $\lambda \approx 2$ mm in GaAs corresponds to a frequency of $\omega_0/2\pi \approx  50$ GHz, yielding an extremely large coupling constant $g_0/2\pi \approx 870 \ \mathrm{MHz}$.   

In atomic systems, the photon decay rate $\kappa$ often limits coherent control.   In the solid state, superconducting transmission line resonators developed for high speed circuitry and photon detection have been produced with Q-factors up to $10^6$ \cite{Resonators};  the photon decay rate $\approx \omega_0/Q$ can thus be very small.   The limiting factor is the charge state dephasing rate $\gamma_c$.  Inelastic transition rates \cite{Fuji02} set a lower bound of a few hundred MHz, and initial observations of coherent charge oscillations reveal a dephasing time near 1 ns, limited by relatively hot (100 mK) electron temperatures \cite{RABI}.   For our calculations, we make the conservative estimate $1/\gamma_c \approx 1$ ns, noting that  the zero temperature value could be an order of magnitude slower.   Regardless of the precise dephasing rate, quantum dot charge states would make rather poor qubits.   Consequently for application to quantum information we must extend the strong charge state coupling to the spin degree of freedom, which decoheres on much longer timescales.

Nevertheless, initial demonstrations of cavity quantum electrodynamics may be possible using only the charge states.   Stimulated emission of photons by a double quantum dot has previously been observed using external microwave  radiation to enhance tunneling rates \cite{DoubleQD}.    If the external source is replaced by the intracavity field of a microwave resonator, stimulated emission can exponentially amplify the field.  For large enough coupling $g_0$, the double dot can potentially act as an on-chip maser.

\subsection{The double dot microscopic maser}

\begin{figure}[htbp]
\vspace{0 in}
\centerline {
\includegraphics[width=3.5in]{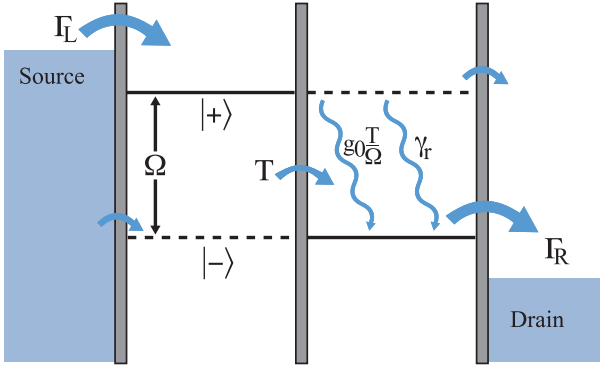}
}
\vspace{0 in}
\caption{\it{The double dot configuration is illustrated in the charge eigenbasis, $|+\rangle, |-\rangle$.  Electrons tunnel from the source into the left dot at rate $\Gamma_L$ and from the right dot to the drain at rate $\Gamma_R$.  For a finite detuning $\Delta$, this pumping can lead to a population inversion.  Decay from the excited $|+\rangle$ state to the ground $|-\rangle$ state occurs via photon emission into the resonator and also via phonon-mediated inelastic decay processes.}} 
\label{setup}
\end{figure}

A double dot operated in the high-bias regime (see Figure~\ref{setup}) can convert electronic potential energy to microwave photons.  By pumping electrons through the double dot, one might hope to induce a population inversion to amplify an applied microwave excitation.  Note that such a device is based on a single emitter, and thus may have properties that differ significantly from those of conventional masers.  In fact, the double dot device corresponds to a direct analogue of the microscopic maser (micro-maser) that has been extensively studied in atomic physics \cite{MandS}.  The micro-maser can be used for unique studies of quantum phenomena including generation of non-classical radiation fields and their non-trivial dynamics \cite{Haroche, Walther}.  

In this paper we will be interested only in the general feasibility of the on-chip micro-maser, and thus we analyze it semiclassically within a rate equation approximation.  In this approximation, effects associated with quantum statistics of the  generated field are disregarded.  Such analysis does however provide a reasonable estimate for the threshold condition and general power.  

The double dot system under consideration has left and right barriers which allow tunneling from the source and to the drain with strength $\Gamma_L$ and $\Gamma_R$ respectively, and one of the dots is capacitively coupled to a resonator as in Figure~\ref{interdot}.   By maintaining a potential energy difference between the two leads, a current is driven through the double dot, and each electron passing through the dot can stimulate emission of a photon into the resonator.   To operate as a maser, however, the double dot must exhibit population inversion, which can only be achieved if electrons preferentially flow in through the excited state and leave via the ground state.   Since finite tunnel coupling $T>0$ is required in order to emit photons into the resonator,  both the excited and ground states $|\pm\rangle$ must be partially delocalized.  This allows electrons to tunnel directly from the source to the ground state and vice versa.   Moreover inelastic decay processes limit how effectively the double dot can convert population inversion into photons.    A careful treatment of pumping and decay rates is therefore needed to demonstrate maser action.

Our semiclassical analysis treats the double dot quantum mechanically, but assumes that the resonator excitations can be described by a coherent state.  We use the density matrix formalism in the rotating wave approximation to derive equations governing the behavior of a double dot coupled to a coherent state of the resonator $\hat{a}|\alpha\rangle = \alpha |\alpha\rangle$.  From Eq.~(\ref{JC}), one can show that the slowly varying components of the density matrix $\hat{\rho}$ in the eigenbasis $\{|+\rangle, |-\rangle\}$  evolve as 
\begin{equation}
\begin{array}{llll}
\dot{\rho}_{++} &=& i \alpha \frac{g_0 T}{\Omega} \left(\rho_{+-} - \rho_{-+}\right)\\
\dot{\rho}_{+-}&=& -i \delta \omega \rho_{+-} +  i \alpha \frac{g_0 T}{\Omega} \left(\rho_{++} - \rho_{--}\right),
\end{array}
\end{equation}
where $\delta\omega = \Omega -\omega_0$, $\dot{\rho}_{--} = -\dot{\rho}_{++}$, and $\dot{\rho}_{-+} = \left(\dot{\rho}_{+-}\right)^*$.  

In addition to this coherent evolution, the density matrix components are affected by dephasing, decay, and coupling to the leads.   In particular, the excited state population $\rho_{++}$ increases at a rate $\Gamma_L |\langle+|L\rangle|^2 (1-\rho_{++} - \rho_{--})$ due to pumping from the source, while it decays at rate $\gamma_r +  \Gamma_R |\langle+|R\rangle|^2$ due to relaxation  to the ground state at rate $\gamma_r$ and loss to the drain.  Similarly, the ground state population $\rho_{--}$ increases at a rate $\Gamma_L |\langle-|L\rangle|^2 (1-\rho_{++} - \rho_{--}) + \gamma_r\rho_{++}$, while losing population at rate $\Gamma_R|\langle-|R\rangle|^2\rho_{--}$.  Note that our density matrix no longer satisfies $\mathrm{Tr}(\rho) = 1$ (since the electron can leave the double dot), and we have accounted for the large charging energy by allowing at most one electron to inhabit the double dot \cite{Stoof}.  

The pumping and decay rates also contribute to reduction of the off-diagonal terms $\rho_{+-}, \rho_{-+}$.  Inelastic decay contributes $\gamma_r/2$ to the charge dephasing rate $\gamma_c$.  The coupling to the leads, however, enters in an asymmetric fashion because $\Gamma_R$ causes direct lifetime broadening, whereas $\Gamma_L$ only affects dephasing through virtual processes which allow the confined electron to scatter off electrons in the source.  We consider the regime $\Gamma_R \approx \Gamma_L$, where these higher order processes can be neglected in comparison to lifetime broadening, and thereby find that the off-diagonal terms decay at a rate $\gamma_{tot} = (\gamma_r + \Gamma_R)/2 + \gamma_c$.   

Taking all terms into account, the density matrix equations of motion are: 
\begin{equation}
\begin{array}{lllll}
\dot{\rho}_{++} &=& i \alpha \frac{g_0 T}{\Omega} \left(\rho_{+-} - \rho_{-+}\right) -\left(\gamma_r + \Gamma_R\frac{\Omega - \Delta}{2 \Omega}\right)\rho_{++}\\
&&+\Gamma_L \frac{\Omega+\Delta}{2\Omega}\left(1-\rho_{++} -\rho_{--}\right)\\\\
\dot{\rho}_{--} &=& i \alpha \frac{g_0 T}{\Omega} \left(\rho_{-+} - \rho_{+-}\right)  -\Gamma_R\frac{\Omega + \Delta}{2 \Omega}\rho_{--}+\gamma_r\rho_{++}\\
&& + \Gamma_L \frac{\Omega-\Delta}{2\Omega}\left(1-\rho_{++} -\rho_{--}\right)\\\\
\dot{\rho}_{+-}&=& -(\gamma_{tot} + i \delta\omega) \rho_{+-} +   i \alpha \frac{g_0 T}{\Omega} \left(\rho_{++} - \rho_{--}\right)\\
\end{array}
\end{equation}
where $\dot{\rho}_{-+} = \left(\dot{\rho}_{+-}\right)^*$.   In the steady state, the time derivatives vanish, and one may easily obtain the polarization $\mathrm{Im}[\rho_{+-}]$ and population inversion $\rho_{++} -\rho_{--}$.  

Simulated emission processes increase the intracavity field.  The amplitude of the coherent state, $\alpha$, grows as $(g_0 T/\Omega)\mathrm{Im}[\rho_{+-}]$ while decaying at rate $\kappa \alpha$ due to the finite linewidth of the resonator.   Expressed in terms of the effective emission rate, $G = (g_0 T/\Omega)^2\gamma_{tot}/(\gamma_{tot}^2 + \delta\omega^2)$, the field growth rate is $\dot{\alpha} = \alpha\left(G \left(\rho_{++} - \rho_{--}\right) - \kappa\right)$.   Substituting the steady state population inversion,  we find the evolution equation for the microwave field:
\begin{equation}
\begin{array}{llll}
\dot{\alpha} &=& \left(\frac{2G\Omega\Gamma_L(\Delta\Gamma_R - \gamma_r \Omega)}{\left(\Gamma_R + 2\Gamma_L\right)\left(\Omega^2 (4 G \alpha^2 + \gamma_r) + 2 T^2 \Gamma_R \right)+ \Delta \Gamma_R\left(2 \Delta \Gamma_L + \Omega \gamma_r\right)}\right) \alpha \\ && -\kappa \alpha.
\end{array}
\label{alph}
\end{equation}
This expression allows derivation of the threshold condition for maser operation, which corresponds to the requirement that the initial growth rate be greater than zero $\left(\dot{\alpha}/\alpha\right)|_{\alpha = 0} > 0$.  Due to saturation effects from the $\alpha^2$ in the denominator of Eq.~(\ref{alph}), the field $\alpha$ grows until $\dot{\alpha} = 0$.  This steady state solution $\alpha_{ss}$ determines the number of generated photons, $\alpha_{ss}^2$, which we identify as the double dot maser figure of merit because it quantifies the amplitude of the microwave field attained in the resonator.    
 
\begin{figure}[htbp]
\vspace{0 in}
\centerline {
\includegraphics[width=3.5 in]{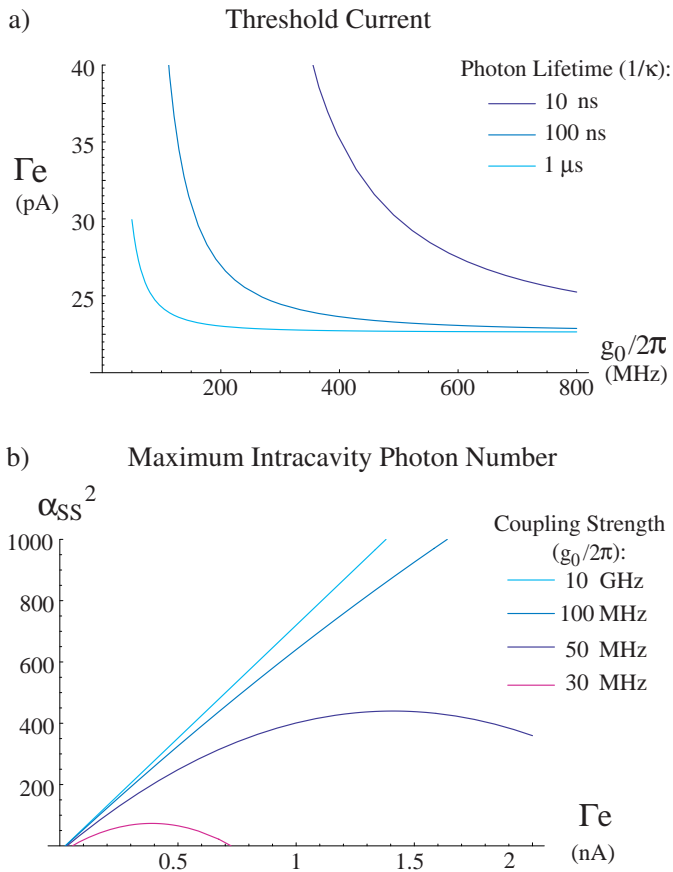}
}
\vspace{0 in}
\caption{\it{a) The threshold  pumping current $e\Gamma$ required for maser operation is shown as a function of dot-resonator coupling strength for three values of $\kappa$, the rate at which photons leak out of the resonator.   b) The maximum number of photons  produced in the resonator as a function of the pumping current  $e\Gamma$ shows saturation due to $\Gamma$-induced dephasing.  Note that the required coupling strength is low: $g_0/2\pi \sim 100$ MHz.  The relevant parameters for these calculations are:  $\Gamma_L = \Gamma_R = \Gamma$, $\Delta/T = 2, \delta\omega = 0,1/ \gamma_c = 1$ ns, $1/\gamma_r = 10$ ns, and in part b) $1/\kappa = 1 \ \mu$s.}} 
\label{maser}
\end{figure}   

The double dot maser cannot be made arbitrarily powerful by increasing the pumping power.   Although increasing $\Gamma_{L,R}$ pumps more electrons through the double dot, opening up conduction to the leads speeds up the dephasing rate, which decreases the effective emission rate $G$.    Consequently there is an optimal pumping rate which maximizes the steady state intracavity field.    

To demonstrate the feasibility of building a double dot maser, we calculate the threshold current and maximum field $\alpha_{ss}^2$ for a realistic set of parameters (see Figure~\ref{maser}).   The threshold condition $\dot{\alpha} > 0$ can be satisfied even for a dot-resonator coupling rate of only $g_0/2\pi = 30$ MHz.    For the calculations shown in Figure~\ref{maser}, we take $\Gamma_L = \Gamma_R = \Gamma$, and moderate values of $\Gamma$ and $g_0$ yield typical resonator excitations of $|\alpha_{ss}|^2 \approx 1000$ photons.  A weak coupling to a nearby transmission line can leak these excitations at a rate $1/\kappa \sim 1 \ \mu$s without significantly affecting the Q of the resonator, allowing emission of around $10^{9}$ photons per second.   The resulting power at $30$ GHz is around 20 fW.  Although these fields are small, they could be detected by photon-assisted tunneling in another mesoscopic two-level system.  For example, recent experiments on a superconductor-insulator-superconductor junction have shown a sub-fW sensitivity to microwave excitations at 25 GHz \cite{DeBlock}.   In conjunction with such sensitive microwave detectors, this on-chip coherent microwave source could provide a useful tool for high frequency spectroscopy of mesoscopic systems.  

\section{Cavity QED with Spin States}

The spin state of an electron in a quantum dot has been suggested as a potential solid state qubit because it possesses good coherence properties \cite{DiVi98}.  While a quantitative value for the spin dephasing rate $\gamma_s$ is unknown, a number of experiments indicate that it is significantly smaller than the charge dephasing rate.   Experiments in bulk 2DEG find $1/\gamma_s = 100$ns \cite{GLoss}, though the situation will be different for confined electrons.  Spin relaxation times of over 50 $\mu$s \cite{Leiven02, Fuji02} indicate that the spin of an electron in a quantum dot can be well protected from its environment.  

The spin does not couple directly to electromagnetic excitations in a resonator, but an indirect interaction is possible by entangling spin with charge.  This technique is similar to Raman transitions in atomic systems, where long-lived hyperfine states interact via an intermediate short-lived excited state.   In particular, by employing the analogue of a Raman transition, which only virtually populates charge state superpositions, we show below that quantum information can be transferred between the stable spin and photon states.

\subsection{Three level systems in double dots}

\begin{figure}[htbp]
\vspace{0 in}
\centerline {
\includegraphics[width=3.25 in]{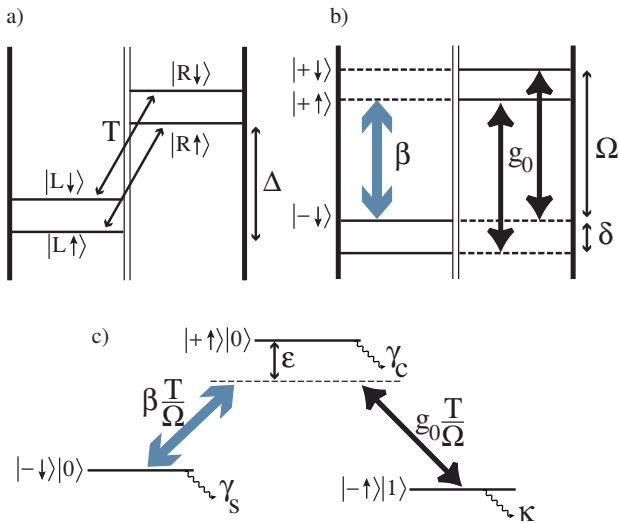}
}
\vspace{0 in}
\caption{\it{ a) An isolated double quantum dot has charge states detuned by $\Delta$ and tunnel coupled by $T$; the spin states are split by $\delta$.  b) In the double dot eigenbasis, an ESR pulse $\beta$ applied locally to the left dot couples $|+\uparrow\rangle$ and $|-\downarrow\rangle$, while the resonator allows transitions between charge states accompanied by emission or absorption of a photon.  c) The three level system has effective coupling strengths $\beta \frac{T}{\Omega}$ and $g_0 \frac{T}{\Omega}$ and losses $\gamma_c, \gamma_s,$ and $\kappa$ due to charge, spin, and photon decoherence respectively.   ESR and resonator frequencies $\nu = \Omega - \delta - \epsilon$ and $\omega_0 = \Omega - \epsilon$ respectively  allow a resonant  two photon transition between the spin levels.}} 
\label{spin}
\end{figure}

A technique based on Raman transitions requires a closed three level system incorporating both the charge and spin degrees of freedom (see Figure~\ref{spin}).   We define the spin states  $|\uparrow\rangle$ and $|\downarrow\rangle$ by applying a static in-plane magnetic field $B_z$, which splits them in energy by $\delta = g \mu_B B_z$ ($ 6.2 \ \mathrm{GHz/Tesla}$ for GaAs).   The electron state is then represented by its charge and spin states $|\pm\rangle \otimes  |\uparrow \downarrow\rangle$ which we abbreviate to $|\pm \uparrow \downarrow\rangle$.   Since an electron in the orbital ground state $|-\rangle$ maintains its spin coherence over long times $\approx 1/\gamma_s$,  we choose $|-\downarrow\rangle$ and $|-\uparrow\rangle$ as the two metastable states.   The resonator couples $|- \uparrow\rangle$ to $|+\uparrow\rangle$; to complete the Raman transition, we need a mechanism which simultaneously flips the charge and spin state between $|+\uparrow\rangle$ and $|-\downarrow\rangle$.   This can be accomplished with a local ESR (electron spin resonance) pulse $2\beta(t) \cos{\nu t}$ acting only on electrons in the left dot.  By tuning the ESR carrier frequency $\nu$  close to the appropriate transition, $|- \downarrow\rangle \rightarrow |+\uparrow\rangle$, we need only consider the near-resonant terms of the local ESR Hamiltonian: 
\begin{equation}
\hat{H}_{ESR} = \beta(t)\frac{T}{\Omega} \big(|+\uparrow\rangle\langle-\downarrow| e^{-i\nu t}+|-\downarrow\rangle\langle+\uparrow| e^{i\nu t}\big).
\label{ESR}
\end{equation}
Choosing the ESR detuning $\Omega - \delta - \nu = \epsilon$ to match the resonator detuning $\Omega - \omega_0 = \epsilon$, we implement a far-off resonant Raman transition. 

Several conditions must be met, however, in order to neglect the energy non-conserving processes as we did in Eq.~(\ref{ESR}).   In particular, the ESR field must be sufficiently weak and sufficiently far detuned from resonance to satisfy the following three inequalities:  $\beta \ll |\nu-\delta|$, $\beta \ll |\Omega + \delta - \nu|$, $(T/\Omega)^2 g_0\beta/\epsilon \ll \delta$.   Physically, this means that the transition rates for undesired spin flips (e.g. $|+\uparrow\rangle$ and $|+\downarrow\rangle$) and the rate for Raman transitions between the wrong levels must both be small compared to the energy detuning associated with each process.  By going far off resonance, we also prevent the transition $|-\downarrow\rangle \rightarrow |+\downarrow\rangle$.  Consequently, the system described by Eq.~(\ref{ESR}) and Eq.~(\ref{JC}) is truncated to three states: $\{|-\uparrow, 1\rangle, |-\downarrow,0\rangle, |+\uparrow, 0\rangle\}$,  where the final symbol indicates the number of excitations in the resonator.  We have thus constructed a solid state analogue of the three-level atom of cavity quantum electrodynamics.  

\subsection{State transfer using a far-off-resonant Raman transition}

The primary goal of the coupling mechanism is to allow interactions between spatially separated spin qubits.  Since auxiliary adjacent spin qubits can accomplish conditional dynamics \cite{DiVi98}, long-distance state transfer is sufficient to attain this objective.  In this section we present an analysis of the coupling rates and the loss of coherence associated with state transfer using a far-off-resonant Raman transition.   

The two states $|-\uparrow,1\rangle$ and $|-\downarrow,0\rangle$ play the role of metastable atomic states, which are coupled via $|+\downarrow, 0\rangle$, which acts as an intermediate excited state.  For constant $g_0 $ and $\beta$ and large detuning $\epsilon$, the metastable states are coupled at an effective transition rate $\chi = (T/\Omega)^2 g_0 \beta/\epsilon$, and state transfer is achieved by pulsing $T/\Omega$ for a time $\tau = \pi/(2 \chi)$.  

Loss of coherence arises both from virtual population of the intermediate state and from dephasing of the spin and photon states.   Virtual population of charge superpositions induces decoherence at a rate $\gamma_{\mathrm{eff}} \approx (T/\Omega)^2\left(P_{\downarrow}\beta^2 + P_{\uparrow} g_0^2\right)/\epsilon^2$ where $P_{\downarrow\uparrow}$ denotes the probability that the system is in the corresponding spin state.  For $\beta$ and $g_0$ of similar magnitude we approximate this decoherence rate by a time independent value $(T/\Omega)^2\langle\beta^2 + g_0^2\rangle/2\epsilon^2$.   The metastable states decohere at rate $\gamma_D$, which is always less than the greater of $\gamma_s$ and $\kappa$ (strictly speaking, $\gamma_D$ also depends on $P_{\uparrow\downarrow}$,  but the dependence is weak since $\gamma_s \approx \kappa$).  During the time required for state transfer, $\tau \approx \pi/2\chi$,  the error probability is $p_{\mathrm{error}} \approx \tau(\gamma_{\mathrm{eff}} +\gamma_D)$.   At optimal detuning \begin{equation}
\epsilon \approx T/\Omega\sqrt{\gamma_c(\beta^2 + g_0^2)/2\gamma_D},
\end{equation}
the  probability of error becomes
\begin{equation}
p_{\mathrm{error}} \approx \sqrt{\gamma_c \gamma_D} \frac{\pi \Omega}{g_0 T }\sqrt{\frac{\beta^2 + g_0^2}{2\beta^2}}.
\end{equation}

As a quantitative example of this technique, suppose $\Omega/2\pi = 2 T /2\pi = 50 \ \mathrm{GHz}$, $g_0/2\pi = 870 \ \mathrm{MHz}$, and $1/\gamma_D$ is set by the expected spin dephasing time $ \approx 1 \mu$s (requiring a resonator quality factor $Q > 5 \cdot 10^4$).  Using an ESR field $\beta /2\pi$ of 1 GHz, the far-off-resonant Raman transition accomplishes state transfer in around 100 ns with $p_{error} \sim 0.2$ for a detuning of $\epsilon/2\pi \approx 15$ GHz.   These numbers easily satisfy the conditions on $\beta$ which allow neglect of energy non-conserving processes.  A high in-plane magnetic field $B_z$ is not needed, since $\nu, \epsilon \gg \beta$, and $\chi/2\pi=15$ MHz, so we merely require $B_z \gg 15$ mT (a low magnetic field is desirable because it presents fewer complications for the superconducting resonator design).  We note that this example provides only a very rough estimate of expected error rates since they depends sensitively on the quantity of interest,  the pulsing mechanism, and the values of $\gamma_s$ and $\gamma_c$.   Optimization of these variables could likely lead to significant improvements in fidelity.

\subsection{Experimental Considerations}
The proposed system fits into ongoing experimental efforts toward single spin initialization and readout \cite{Leiven02}.  Local ESR,  however,  has not yet been experimentally demonstrated, and will likely represent the most challenging element in our proposal.  Nevertheless, a variety of tools are now being developed which may permit local spin manipulation.  Consequently we now consider several strategies for achieving the simultaneous charge transition and spin flip required for our scheme.  

One promising route to spin resonance is g-factor engineering \cite{Awschalom2001}.    In our proposal, even a static ESR interaction ($\nu = 0$) could be used to flip spins, provided that $\delta \gg \beta$.   Modulation of one component of the g-factor tensor, e.g. $g_{xx}$, could thus turn on and off the effect of a static applied field $B_x$ by modulating the Zeeman term $g_{xx} \mu_B B_x \sigma_x$ in the Hamiltonian.  By making the electrodes small enough, this g-factor shift could be applied to a single dot.  Although working near $\nu = 0$ will vastly reduce the heat load associated with ESR, the system will be more sensitive to low frequency fluctuations in the electromagnetic environment.  

Alternately, high frequency anisotropic g-factor modulation can induce spin flips using a static magnetic field and microwave electric fields \cite{Awschalom2003}.   As in the case of static g-factor engineering, the electric field could be applied locally.  The resulting ESR coupling strength depends on the voltage-induced change in anisotropic g-factor $\Delta g_{xz}$ multiplied by the applied static field.   If one can engineer $\Delta g_{xz}= 0.03$ (static g-factor engineering can induce $\Delta g_{zz} = 0.16$ \cite{Awschalom2001}), a static magnetic field of $16$~T would produce the desired local ESR strength.   Operation at such large fields, however, would necessitate using a type II superconductor to construct the resonator.  Since the flux-pinning mechanisms which allow high-field superconductivity also contribute to residual surface impedance, resonators constructed of type II materials would likely have lower quality factors. 

Other strategies combine a global ESR pulse with some other mechanism that couples spin and charge, for example a spin-dependent tunneling rate.  In this case, the charge eigenstates can be made different for the two spins, so that $\langle -_{\downarrow}| +_{\uparrow}\rangle  = \eta \neq 0$ (here $|\pm_{\sigma}\rangle$ is the charge eigenbasis for an electron with spin $\sigma$).   If the global ESR strength is $\beta_{global}$, the coupling between $|+\uparrow\rangle$ and $|-\downarrow\rangle$ is $\eta \beta_{global}$ whereas the additional decoherence due to the slightly different charge distributions of $|-\uparrow\rangle$ and $|-\downarrow\rangle$ is only $\eta^2 \gamma_c$.   Sufficiently large $\beta_{global}$ and small $\eta$ permit state transfer with negligible contribution to dephasing.

\section{Control of Low-Frequency Dephasing with a Resonator}

Thus far we have assumed a single-mode resonator and included only energy-conserving processes.   A more careful analysis incorporates the energy non-conserving terms of Eq.~(\ref{SB}), which lead to  corrections scaling as $g_{x,z}^2/\omega_n^2$.   If the resonator has a minimum frequency $\omega_0\gg g_0$, these terms are small and can be neglected.   However, as the minimum frequency decreases, energy non-conserving terms may become important.

\begin{figure}[htbp]
\vspace{0 in}
\centerline {
\includegraphics[width=3.25in]{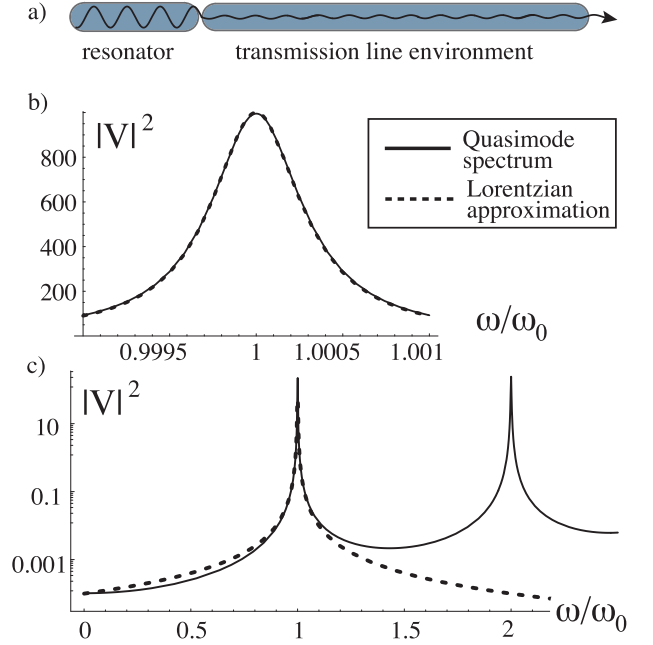}
}
\vspace{0 in}
\caption{\it{a) Losses in the resonator may be modeled by a weak coupling to a semi-infinite transmission line.  b)  For a high-Q resonator, the quasimode voltage spectrum at $x = -l$ (solid)  is well approximated near the fundamental mode by a Lorentzian (dashed) with half width $\omega_0/Q$.  c) A logarithmic plot of the same fit shows that the Lorentzian approximation provides an upper bound on the effects of dephasing due to low-frequency modes.}} 
\label{quasimode}
\end{figure}

In an experimental implementation a major concern could be that  coupling a double dot to a macroscopic resonator (which is in turn coupled to the environment) might drastically increase the charge decoherence rate.  Consequently we examine the resonator modes more rigorously.   Following \cite{Quasi}, we model the environment as a transmission line of length $L\rightarrow \infty$ capacitively coupled to the resonator, and diagonalize the resonator + transmission line system (see Figure \ref{quasimode}a).  The discrete mode operators $\hat{a}_n, \hat{a}_n^{\dagger}$ are replaced by the  creation and annihilation operators $\hat{A}_k, \hat{A}_k^{\dagger}$ for the eigenmodes of the total infinite system, which have a continuous spectrum of modes. Since arbitrarily low frequencies are represented, the corrections $\propto (g^{(k)}_{x,z}/\omega_k)^2$ may no longer be negligible.

If the double dot were coupled directly to a transmission line, the low-frequency dephasing would indeed pose a problem.  However, the effect is mitigated because  the resonator is only sensitive to environmental noise within $\omega_0/Q$ of the resonant frequency.  For high quality factors, the resonator voltage spectrum at quasimode frequencies near and below the fundamental mode approaches a Lorentzian (see Figure~\ref{quasimode}),
\begin{equation}
|V_k|^2 \approx \frac{\hbar \omega_k}{2 C_0 L}\frac{Q\omega_0^2}{Q^2(\omega_k-\omega_0)^2 + \omega_0^2},
\end{equation}
which vanishes as $\omega_k \rightarrow 0$.  The coupling strength $g^{(k)}_{x,z}$ between the quantum dot and the quasimode is proportional to this voltage, so the problematic coupling to low frequency modes is strongly suppressed.  

To illustrate the suppression of decoherence, consider the dephasing effect of the term $\sum_k{g^{(k)}_z (\hat{A}_k + \hat{A}_k^{\dagger}})\Sigma_z $ in the Hamiltonian of Eq.~(\ref{SB}) in two situations:  (a) the double dot is coupled directly to the environment, so $A_k$ annihilates a mode of a semi-infinite transmission line, and (b) the double dot is coupled to a resonator which is in turn coupled to a transmission line environment, so $A_k$ annihilates a quasimode of the resonator+transmission line system.   To calculate dephasing, we start in the vacuum state for electromagnetic degrees of freedom and require that the coupling $g_z^{(k)}$ is gradually turned on and off with time dependence $f(t) = \int_{-\infty}^{\infty}e^{i \omega t}\tilde{f}(\omega)d\omega$.  After tracing over the transmission line degrees of freedom,  the component $|+\rangle\langle-|$ of the reduced charge state density matrix has decreased by $e^{-2\kappa_{a}}$ in case (a) and $e^{-2\kappa_{b}}$ in case (b) where   
\begin{eqnarray}
\kappa_a &=& \int_0^{\infty}\frac{L Z_0 C_0}{\pi}|g_0 \tilde{f}(\omega)|^2 d\omega \\
\kappa_b &=& \ \int_0^{\infty}\frac{L Z_0 C_0}{\pi}\frac{Q\omega_0^2 |g_0 \tilde{f}(\omega)|^2}{Q^2(\omega-\omega_0)^2 + \omega_0^2} d\omega.
\end{eqnarray}
Assuming a slowly varying $f(t)$ and a high-Q cavity, examination of Eq. (15,16) directly yields 
\begin{equation}
\frac{\kappa_b}{\kappa_a} \approx \frac{1}{Q}
\end{equation}
i.e. the low-frequency dephasing of the qubit is greatly suppressed.  
This result indicates that in situations where the double dot dephasing is dominated by coupling to electromagnetic modes of the transmission line, a resonator provides good protection from low-frequency dephasing.
 
 \section{Conclusion}
 
In this paper we describe how effects familiar from atomic cavity quantum electrodynamics may be observed in highly tunable solid state devices.  Quantum dot charge states are limited by a fast decoherence rate, but a solid state maser may nonetheless demonstrate coherent interactions between a double dot and a resonator.   For quantum information systems, lower decoherence rates are required, and we have illustrated how a long-range interaction between long-lived spin states may be implemented with only virtual population of intermediate charge states.   This system represents an opportunity to manipulate electron spins and charges with a level of control usually associated with atomic physics, and illustrates how techniques pioneered quantum optics can find application in a solid state context.  

On a broader level, our work also demonstrates how a high-quality resonator can serve as a quantum coherent data bus between qubits.   Such a data bus could provide an interaction between different types of quantum systems.   Cooper pair boxes \cite{JJRes} or even Rydberg atoms \cite{Anders} could thereby interact with electron spins.   If sufficiently strong coupling mechanisms can be found, these artificial atoms and microwave resonators could have an important role to play as tunable, integrable, and scalable coherent quantum systems. 
 
 \section{Acknowledgements}
 
The authors wish to thank Steve Girvin, Rob Schoelkopf, Caspar Van der Waal, Charles Marcus, Leo Kouwenhoven, and Leiven Vandersypen for valuable discussions.  This work is supported by the NSF, ARO, and David and Lucille Packard  and Alfred Sloan Foundations.  L.C. acknowledges support from the Hertz Foundation, and A.S. is supported by the Danish Natural Science Research Council and by the Natural Science Foundation through its grant to ITAMP.

\vspace{0.1in}
\small{

}

\end{document}